\documentclass[showpacs,prc,amsmath,amssymb,twocolumn]{revtex4}

\usepackage{graphicx}% Include figure files
\usepackage{dcolumn}% Align table columns on decimal point
\usepackage{bm}% bold math

\begin{document}

%\preprint{APS/123-QED}

\title{Dynamical and sequential decay effects on isoscaling and density dependence of the symmetry energy}

\author{W. D. Tian$\,^{1}$\footnote{E-mail: tianwendong@sinap.ac.cn}, Y. G. Ma$\,^{1}$\footnote{E-mail:
ygma@sinap.ac.cn}, X. Z. Cai$\,^{1}$, D. Q. Fang$\,^{1}$, W.
Guo$\,^{1,2}$, C. W. Ma$\,^{1,2}$, G. H. Liu$\,^{1,2}$, W. Q.
Shen$\,^{1}$, Y. Shi$\,^{1,2}$, H. W. Wang$\,^{1}$, K.
Wang$\,^{1,2}$, W. Xu$\,^{1}$, T. Z. Yan$\,^{1,2}$}
\address{
1) Shanghai Institute of Applied Physics, Chinese Academy of
Sciences, P. O. Box 800-204, Shanghai 201800, China \\
2) Graduate School of Chinese Academy of Sciences, Beijing,
100049, China}

\date{\today}

\begin{abstract}
The isoscaling properties of the primary and final products are
studied via isospin dependent quantum molecular dynamics
 (IQMD) model and the followed sequential decay model GEMINI, respectively. It is found that the
 isoscaling parameters $\alpha$ of both primary and final products
 keep no significant change for light fragments, but increases
 with the mass for intermediate and heavy products. The
dynamical effects on isoscaling are exhibited by that $\alpha$
value decreases a little with the evolution time of the system,
and opposite trend for the heavy products. The secondary
 decay effects on isoscaling are reflected in the increasing of the $\alpha$ value
 for the final products which experiences secondary decay process.
 Furthermore the density dependence of the symmetry energy has
 also been explored, it is observed that in the low densities the
 symmetry energy coefficient has the form of
$C_{sym}(\rho)\sim C_{0}(\rho/\rho_{0})^{\gamma}$, where $\gamma =
0.7 \sim 1.3$ for both primary and final products, but $C_{0}$
have different values for primary and final products. It is also
suggested that it might be more reasonable to describe the density
dependence of the symmetry energy coefficient by the
$C_{sym}(\rho/\rho_{0})\approx
C_{1}(\rho/\rho_{0})^{\gamma_{soft}} +
C_{2}(\rho/\rho_{0})^{\gamma_{stiff}}$ with $\gamma_{soft}\leq 1$,
$\gamma_{stiff}\geq 1$ and $C_{1}, C_{2}$ constant parameters.
\end{abstract}

\pacs{24.10.Nz, 24.10.-i, 25.70.Pq, 25.70.-z, 24.80.+y}

\maketitle

\section{\label{sec:sec1}Introduction}

The isotopic composition of the nuclear reaction products
\cite{Li01} contains important information on the role of the
isospin on the reaction process. Recently increased interest in
the $N/Z$ degree of freedom and its equilibration, as well as the
isospin asymmetry dependent terms of the nuclear equation of state
(EOS) \cite{To99, Li98, Ma99,Mu95, Bo94, Gu01}, has motivated
detailed measurements of the isotopic distributions of reaction
products. It has been shown that isospin effects can be studied by
comparing the yields of fragments from two similar reactions that
differ only in the isospin asymmetry \cite{Xu00,Ts01}, in this
case, the effect of sequential decay of primary fragments can be
bypassed to a large extent. It has been revealed that for
statistical fragment production mechanism(s), if two reactions
occurring at the similar temperature have different isospin
asymmetry, the ratio $R_{21}(N,Z)$ of the yields of a given
fragment $N$ and $Z$ obtained from the two reactions 2 and 1
exhibits an exponential dependence on $N$ and $Z$ of the form
\cite{Ts01,Ts01a,Ts01b}
\begin{equation}
R_{21}(N,Z) = Y_2(N,Z)/Y_1(N,Z) = C \exp(\alpha N+\beta Z),
\label{eq:one}
\end{equation}
where $\alpha$ and $\beta$ are two scaling parameters and $C$ is
an overall normalization constant. This behavior is called
isoscaling \cite{Ts01a}. Isoscaling has been obtained in different
reactions and theoretical calculations, such as in evaporation
reactions \cite{Br93}, deep inelastic reactions \cite{Vo78,So03},
fission reactions \cite{Ve04} and multifragmentations
\cite{Xu00,Ts01,Bo02,Ge04,Sh04}, where is the biggest amount of
experimental isoscaling data comes from. Isoscaling phenomena in
different reaction mechanisms have also been investigated by
various theoretical models, such as Langevin equation combining
with the statistical decay model for the fission dynamics
\cite{Wa05,Ma05}, isospin-dependent Lattice Gas model \cite{Ma04}
for the multifragmentations, Anti-symmetrized Molecular Dynamics
(AMD) model \cite{On03}, Isospin dependent Quantum Molecular
Dynamics model \cite{Ti05}, Microcanonical Statistical
Multifragmentation model, Expanding Emitting Source model,
Canonical model \cite{Ts01b}, Microcanonical Multifragmentation
model \cite{Ra05} and so on.

The isospin dependence of the nuclear equation of state (EOS) is
one of the most important properties in nuclear matter and
reactions, in particular in nuclear astrophysics, supernova and
neutron stars \cite{Li01,La00}. Although the nuclear symmetry
energy at normal nuclear matter density $\rho_{0} = 0.16 fm^{-3}$
has been determined to be around 30 MeV from the empirical
liquid-drop mass formula \cite{My66}, however, its values at sub-
and super-densities are poorly known. Studies based on various
theoretical models also give widely different predictions
\cite{Li01, Sh04, Fe05, Ch05}. Multifragmentation is generally
considered as a low density phenomenon with a high degree of
thermalization which is believed to be reached. It has been shown
\cite{Ts01b,Bo02} that the isoscaling parameter $\alpha$ is
directly related to the coefficient $C_{sym}$ of the symmetry
energy term of the nuclear binding energy, the following relation
has been obtained both in the framework of the grand-canonical
limit of the statistical multifragmentation model \cite{Ts01b} and
in the expanding-emitting source model \cite{Bo02}:

\begin{equation}
\alpha =
4\frac{C_{sym}}{T}\Big[\Big(\frac{Z_{1}}{A_{1}}\Big)^2-\Big(\frac{Z_{2}}{A_{2}}\Big)^2\Big]
\label{eq:two}
\end{equation}
where $Z_{1}$, $A_{1}$, and $Z_{2}$, $A_{2}$ refer to the charge
number and mass number of the fragments from reactions 1 and 2
respectively. Using this relation, from the extracted values of
$\alpha$ and $T$ of the fragments in the reaction, symmetry energy
coefficient $C_{sym}$ in the EOS could be derived.

The paper aims to investigate isoscaling in multifragmentation
reactions by the dynamical IQMD model and the afterburner
sequential decay model GEMINI calculation, explore the dynamical
and secondary decay effects on the isoscaling phenomenon.
Different isoscaling properties between light and heavy products
have revealed. Through the relationship of equation (\ref{eq:two})
between isoscaling parameter $\alpha$ and symmetry energy
coefficient $C_{sym}$, the form of symmetry energy density
dependence can be deduced. Section \ref{sec:sec2} gives a brief
review of the IQMD and GEMINI model. Section \ref{sec:sec3}
presents the IQMD and IQMD+GEMINI calculation results on
isoscaling obtained from the different system evolution time,
primary and final products, discusses the properties of the
isoscaling parameter $\alpha$. Section \ref{sec:sec4} investigates
the density dependence of the symmetry energy coefficient
$C_{sym}$ for the primary and final products following some
results in section \ref{sec:sec3}. The conclusions are drawn in
section \ref{sec:sec5}.

\section{\label{sec:sec2}Model overview}

\subsection{Dynamical model: IQMD}

The Quantum Molecular Dynamics (QMD) model approach is an $n$-body
theory to describe heavy ion reactions from intermediate energy to
several GeV/nucleon, it is classical in essence because the time
evolution of the system is determined by classical canonical
equation of motion, however many important quantum features are
included in this prescription. At intermediate energies the heavy
ion collision is mainly governed by three components: the mean
field, two-body collision and Pauli blocking. General review about
the QMD can be found in reference of Aichelin \cite{Ai88-91}
together with several other similar models such as AMD \cite{On92}
and Constrained Molecular Dynamics (CoMD) model \cite{Pa01}.

The IQMD model is based on the general QMD to include explicitly
isospin-degrees of freedom \cite{Ch98}. In the QMD model, each
nucleon is represented by a Gaussian wave packet,
\begin{equation}
\psi_i(\textbf{r},\textbf{p})=\frac{1}{(2\pi
L)^{3/4}}\exp\Big[-\frac{(\textbf{r}-\textbf{r}_i)^{2}}{4L}+i\textbf{r}\cdot\textbf{p}\Big]
\label{eq:three}
\end{equation}
where $\bf{r}_i$ is the center of the $i$th wave packet in the
coordinate space, and $L$ is the so-called Gaussian wave packet
width (here $L$ = 2.16${fm}$$^2$). The total $N$-body wave
function is assumed to be the direct product of these coherent
states. Through a Wigner transformation of the wave function,
defining $L = \sigma_{r}^{2}$ and $\sigma_{p}^{2} =
\hbar^{2}/4\sigma_{r}^{2} = \hbar^{2}/4L$ the $N$-body phase-space
distribution function is given by a symmetry form in coordinate
and momentum space
\begin{equation}
f(\textbf{r},\textbf{p})=\sum_i{\frac{1}{(\pi\hbar)^3}
\exp\Big[-\frac{(\textbf{r}-\textbf{r}_i)^2}{2\sigma_{r}^{2}}-
\frac{(\textbf{p}-\textbf{p}_i)^2}{2\sigma_{p}^{2}}\Big]},
\label{eq:four}
\end{equation}
here $\bf{r}_i$ and $\bf{p}_i$ are the center of the $i$th wave
packet in the coordinate and momentum space, $\sigma_{r}$ and
$\sigma_{p}$ are the widths of wave packets in coordinate and
momentum space, respectively. The Wigner representation of the
Gaussian wave packet obeys the uncertainty relation
$\sigma_{r}\sigma_{p} =  \hbar/2$. The densities in coordinate and
in momentum space are given by

\begin{equation}
\rho(\textbf{r})=\sum_{i}{\frac{1}{(2\pi
\sigma_{r}^{2})^{3/2}}\exp\Big[-\frac{(\textbf{r}-\textbf{r}_i)^2}{2\sigma_{r}^{2}}\Big]},
\label{eq:five}
\end{equation}

\begin{equation}
\rho(\textbf{p})=\sum_{i}{\frac{1}{(2\pi
\sigma_{p}^{2})^{3/2}}\exp\Big[-\frac{(\textbf{p}-\textbf{p}_i)^2}{2\sigma_{p}^{2}}\Big]}.
\label{eq:six}
\end{equation}

In the IQMD model, the nucleons in a system move under a
self-consistently generated mean field, and the time evolution of
$\textbf{r}_i$ and $\textbf{p}_i$ is governed by Hamiltonian
equations of motion

\begin{equation}
\dot{\textbf{r}}_i=\frac{\partial H}{\partial \textbf{p}_i},
\dot{\textbf{p}}_i=-\frac{\partial H}{\partial \textbf{r}_i}.
\label{eq:seven}
\end{equation}

The Hamiltonian $H$ which consists of both kinetic energy and
effective interaction potential energy is given by

\begin{equation}
H =
\sum_i{\frac{\textbf{p}_i^2}{2\mu}}+U^{dd}+U^{yuk}+U^{sym}+U^{coul},
\label{eq:eight}
\end{equation}
where $\mu$ is the mass of a nucleon, $U^{dd}$ the
density-dependent (Skyrme) potential, $U^{yuk}$ the Yukawa
potential, $U^{sym}$ the symmetry energy term and $U^{coul}$ the
coulomb energy, they have the following forms

\begin{equation}
U^{dd} =
\alpha(\frac{\rho}{\rho_0})+\beta(\frac{\rho}{\rho_0})^{\gamma},
\label{eq:nine}
\end{equation}

\begin{eqnarray}
U^{yuk}&=&\frac{v_y}{2}\sum_{i,j\neq i}\frac{1}{r_{ij}}\exp(\sigma_{r}^{2}m^2) \nonumber\\
& &\cdot\Big[\exp(mr_{ij})erfc(m\sigma_{r}-\frac{r_{ij}}{2\sigma_{r}})\nonumber\\
&
&-\exp(mr_{ij})erfc(m\sigma_{r}+\frac{r_{ij}}{2\sigma_{r}})\Big],
\label{eq:ten}
\end{eqnarray}

\begin{equation}
U^{sym}=\frac{C_{sym}}{2\rho_0}\sum_{i,j\neq
i}{\tau_{iz}\tau_{jz}\frac{1}{(4\pi
\sigma_{r}^{2})^{3/2}}\exp\Big[-\frac{(r_i-r_j)^2}{4\sigma_{r}^{2}}\Big]},
\label{eq:eleven}
\end{equation}

\begin{equation}
U^{coul}=\frac{e^2}{4}\sum_{i,j\neq
i}{\frac{1}{r_{ij}}(1+\tau_{iz})(1-\tau_{jz})erfc(\frac{r_{ij}}{4\sigma_{r}^{2}})}.
\label{eq:twelve}
\end{equation}
Here $r_{ij}$ = $|r_i-r_j|$ is the relative distance of nucleon
$i$ and $j$, $\tau_{iz}$ the $z$th component of the isospin degree
of freedom for the $i$th nucleon, which is equal to 1 and -1 for
proton and neutron, respectively, $C_{sym}$ is the symmetry
potential coefficient. The parameters in this work are listed in
Table \ref{tab:table1}.

\begin{table}
\caption{\label{tab:table1}The parameters adopted in present
work.}
\begin{ruledtabular}
\begin{tabular}{cccccccc}
 $\alpha(GeV)$\footnotemark[1] &$\beta(GeV)$ &$\gamma$ &$\rho_0(fm^{-3})$ &$v_y(GeV)$ &$m$ &$C_{sym}(GeV)$\\
 \hline
 -0.356 &0.303 &1.17 &0.16 &-0.0024 &0.83 &0.032 \\
\end{tabular}
\end{ruledtabular}
\footnotemark[1]{Here $\alpha$, $\beta$ and $\gamma$ values
correspond to soft potential}
\end{table}

The nucleon-nucleon ($NN$) cross section used in present IQMD
model is experimental parametrization \cite{Ch68} ($\sigma_{exp}$)
which is isospin dependent, the neutron and proton are
distinguished from each other in the initialization of projectile
and target nuclei, and the Pauli blocking of proton and neutron in
IQMD is also treated separately as well \cite{Ch98}. In our work,
the fragment (or cluster) is recognized by a simple coalescence
model: i.e., nucleons are considered to be part of a cluster in
every moment at least another nucleon is closer than $r_{min}\leq
3.5fm$ in the coordinate space.

\subsection{\label{sec:level2}Sequential decay model: GEMINI}

The primary fragments from IQMD are excited, in order to compare
with the experimental observable, the GEMINI sequential decay
model \cite{Ch88,Gemini} has been used as an afterburner to follow
the de-excitation of these excited fragments. In GEMINI all
possible sequential decay channels from light particle
evaporations to symmetry fission are considered. The sequential
decay is traced by a Monte-Carlo technique until the excitation
energy of the excited fragment is exhausted. In IQMD calculation
the angular momentum is not considered, hence the secondary decay
treatment in the GEMINI does not include the angular momentum
calculation. In other word, for each sequential decay tracing, the
initial angular momentum is set to zero in GEMINI, and in present
work all asymmetric divisions are considered. More details about
GEMINI can be found in Refs.~\cite{Ch68, Gemini}. IQMD calculation
is on the basis of event-by-event, so the secondary decay by
GEMINI is managed event-by-event too. Each fragment calculated in
one IQMD event will put into the GEMINI to let them decay. The
excitation energy of the primary fragments is calculated by
following equation

\begin{equation}
E^{*} = E_{inc}^{c.m.}- \sum_{\text{mult}}E_{kin}^{c.m.} - Q,
\label{eq:thirteen}
\end{equation}
where $E_{inc}^{c.m.}$ is the incident kinetic energy in Center of
Mass (c.m.) system and $E_{kin}^{c.m.}$ is the fragment kinetic
energy in c.m. system. The summer is made over all the fragments
of each event, the $Q$-value is calculated by the mass excess in
each event, then the excitation energy per nucleon $E^{*}/A$ was
uniformly distributed in all nucleons.

In this paper all asymmetric divisions was considered(including
both intermediate mass fragment(IMF) and fission decay) in the
GEMINI secondary decay calculation, after GEMINI calculation, the
final products were collected inclusively again.

\section{\label{sec:sec3}isocaling properties and the dynamical and secondary decay effects}

\begin{figure}
\includegraphics[width=0.5\textwidth]{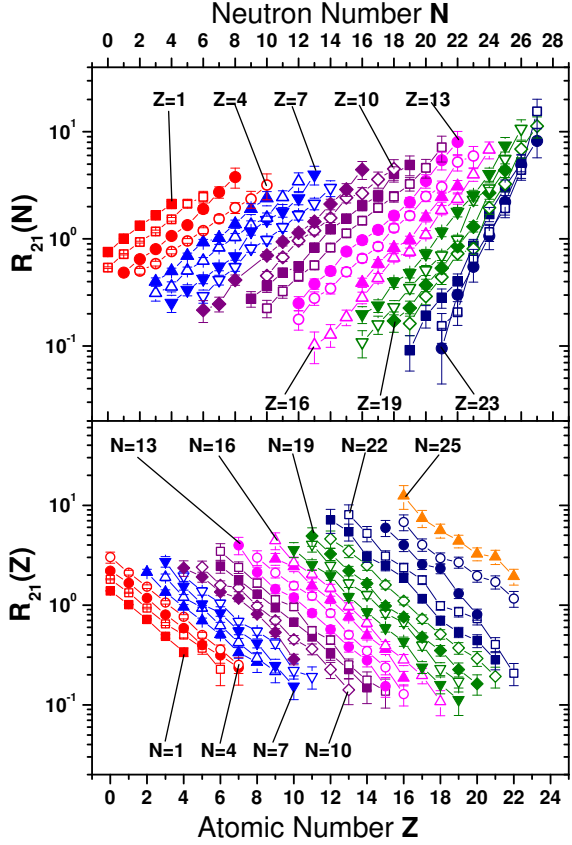}
\caption{\label{fig:fig1} Isotopic yield ratios (upper panel) and
Isotonic yields ratios (bottom panel) of primary fragments at
reaction time $t = 240fm/c$ between $^{48}Ca + ^{48}Ca$ and
$^{40}Ca + ^{40}Ca$ in the IQMD model with impact parameter $b = 1
fm$ and incident energy $E_{inc} = 35 MeV/A$, different symbols
from left to right represent the isotope sequence from $Z = 1$ to
$Z = 23$ (upper panel) and isotone sequence from $N = 1$ to $Z =
25$ (bottom panel), in this figure the error bar comes from only
statistics}
\end{figure}

Calculations were carried out in two similar central collisions
with impact parameter $b = 1$ of $^{40}$Ca + $ ^{40}$Ca and
$^{48}$Ca + $^{48}$Ca at incident energy $E/A = 35$ MeV. In order
to investigate the isoscaling dynamical effect, at different
reaction time fragments were recorded until 400 fm/$c$, about 50
thousands events are simulated.

Equation (\ref{eq:one}) can be expressed as a function of only
atomic number Z when the neutron number N is fixed
\begin{equation}
R_{21}(Z) = Y_2(N,Z)/Y_1(N,Z) = C^{'}\exp(\alpha N),
\label{eq:fourteen}
\end{equation}
or a function of neutron number N when the atomic number Z is
fixed
\begin{equation}
R_{21}(Z)=Y_2(N,Z)/Y_1(N,Z)=C^{''}\exp(\beta Z).
\label{eq:fifteen}
\end{equation}

Isoscaling of hot fragments was observed at different reaction
time from $t$ = 160fm/$c$ to $400$fm/$c$, fragments mentioned here
include all of the fragments formed in that moment. In fig. 1 two
examples of $R_{21}$ were shown for the fragments at reaction time
$t$ = 240 fm/$c$ as a function of atomic number $Z$ and neutron
number $N$, respectively. $R_{21}$ shows perfect linear dependence
as a function of $N$ or $Z$ from light fragments to heavy
fragments. Here only reaction time $t$ = 240 fm/$c$ are plotted as
one sample, the results at other reaction times have similar
behaviors. Here we adopt the widely used convention to denote with
the index "2" the more neutron-rich system $^{48}$Ca + $^{48}$Ca
and with the index "1" the more neutron-poor system $^{40}$Ca +
$^{40}$Ca.

\begin{figure}
\includegraphics[width=0.5\textwidth]{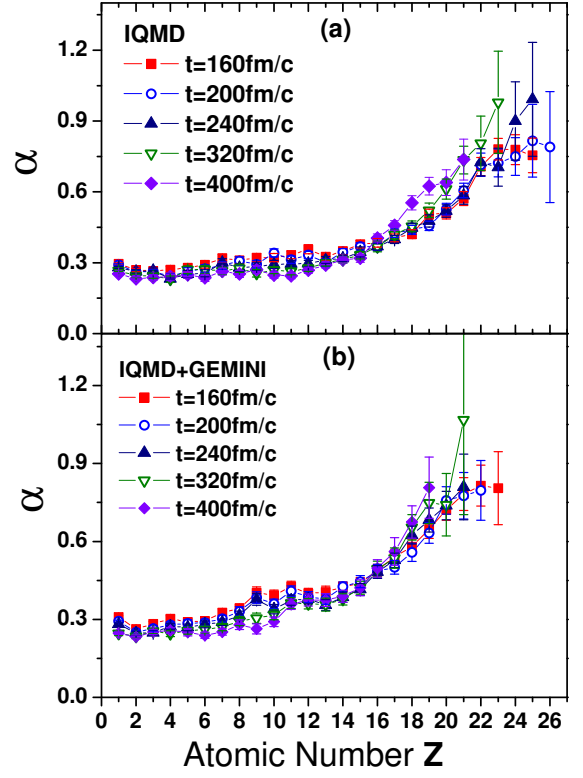}
\caption{\label{fig:fig2} Isoscaling parameter $\alpha$ as a
function of the fragment atomic number $Z$, different symbols
represent $\alpha$ at different reaction time, panel (a): IQMD
calculation results; panel (b): IQMD+GEMINI calculation results.}
\end{figure}

From Equation (\ref{eq:fourteen}) and (\ref{eq:fifteen}) the
isoscaling slope parameters $\alpha$ and $\beta$ can be directly
extracted, in this paper only parameter $\alpha$ is analyzed,
since parameter $\beta$ has similar trend with $\alpha$ except
that $\beta$ value is negative. In fig. \ref{fig:fig2}(a),
parameter $\alpha$ as a function of the fragment charge number $Z$
was printed at several reaction time, one can find that for the
light and intermediate mass fragments, such as $Z \leq 15$,
$\alpha$ value does not change too much with the fragment charge
number $Z$, but for the heavy fragments, $\alpha$ shows different
behavior from the light ones, it increases with the fragments
charge number $Z$. One can also find that the isoscaling parameter
$\alpha$ presents very small decrease when the reaction time
becomes longer, at $t$ = 140 fm/$c$ $\alpha$ lies on the top, $t$
= 400 fm/$c$ $\alpha$ locates at the bottom. However, this is
suitable only for the light products, and  it is opposite for the
heavy products though the dynamical effect is not clear.

\begin{figure}
\includegraphics[width=0.5\textwidth]{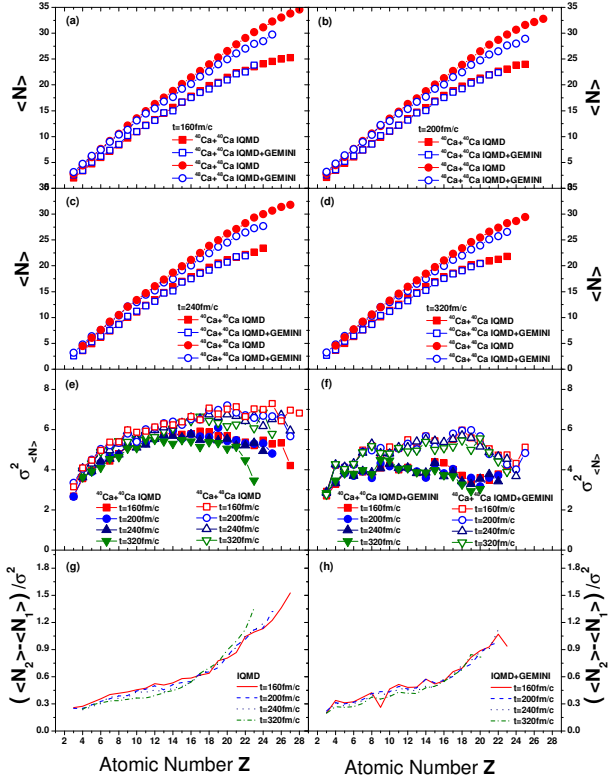}
\caption{\label{fig:fig3} The centroid $\langle N\rangle$ and
width  $\sigma^{2}$ of Gaussian isotopic distribution as a
function of the atomic number $Z$ at different reaction times for
the primary products in IQMD calculation and final products in
IQMD+GEMINI calculation. Panels (a) -(f): centroid $\langle
N\rangle$ as a function of $Z$ at different evolution time in two
systems; panel (e) and (f): $\sigma^{2}$ as a function of $Z$, (g)
and (h): calculated $(\langle N_{2}\rangle-\langle
N_{1}\rangle)/\sigma^{2}$ from upper panels.}
\end{figure}

From a practical point of view, the isotopic distributions of a
given $Z$ fragment $Y(N)|_{Z}$ can be approximately described by
Gaussian function. Using the general notation for the isotopic
distribution in a given $Z$, $Y(N)|_{Z}$ can be described by a
single Gaussian function as

\begin{equation}
Y(N)|_{Z}=C\exp\Big[-\frac{(N- \langle
N_{Z}\rangle)^{2}}{2\sigma_{Z}^{2}}\Big],
\end{equation}
where $\langle N_{Z}\rangle$ is the centroid of isotopic
distributions, and $\sigma_{Z}$ describes the variance of
distributions for each element of charge $Z$, then for a fixed
element of charge $Z$ the ratio of isotopic yields writes

\begin{eqnarray}
R_{21}(N)& = & \frac{Y_{2}(N)|_{Z}}{Y_{1}(N)|_{Z}} \nonumber\\
&&=C^{'}\exp\Big[N\Big(\frac{\langle
N_{2}\rangle}{2\sigma_{2}^{2}}-\frac{\langle
N_{1}\rangle}{2\sigma_{1}^{2}}\Big)\Big] \nonumber\\
&&\cdot\exp\Big[N^{2}\big(\frac{1}{2\sigma_{1}^{2}}-\frac{1}{2\sigma_{2}^{2}}\Big)\Big]
\label{eq:seventeen}
\end{eqnarray}
where $\langle N_{1}\rangle$ and  $\langle N_{2}\rangle$ are the
centroids for isotopic distributions of neutron-poor system
$^{40}$Ca + $^{40}$Ca and neutron-rich system $^{48}$Ca +
$^{48}$Ca, $\sigma_{1}$ and $\sigma_{2}$ describe the variance of
distributions in these two systems, respectively, here the
fragment charge number $Z$ is fixed.

From equation (\ref{eq:seventeen}) one can notice that if the
second order $N^{2}$ in the exponential function can be neglected,
$\ln(R_{21}(N))$ is linear dependence on neutron number $N$ for a
fixed element with charge number $Z$, and $\alpha$ value can be
expressed as $\langle N_{2}\rangle/2\sigma_{2}^{2}-\langle
N_{1}\rangle/2\sigma_{1}^{2}$. Fig. \ref{fig:fig3} shows the
centroid $\langle N\rangle$  and width $\sigma^{2}$ of Gaussian
isotopic distributions as a function of atomic number $Z$ of the
emitted products at different reaction time ( panel (a) - (f))
before and after secondary decay, panel (a) to (d), the difference
of $\langle N_{2}\rangle$ in $^{48}$Ca + $^{48}$Ca system and
$\langle N_{1}\rangle$ in $^{40}$Ca + $^{40}$Ca system increases
when atomic number $Z$ of the emitted products becomes large, in
panel (e) and (f) of fig. \ref{fig:fig3}, the  width $\sigma^{2}$
of isotopic distributions does not same in two similar reactions
too, it increases with the increasing of fragment atomic number
$Z$, reaches saturation at some point. In fig. \ref{fig:fig3}(g),
$(\langle N_{2}\rangle-\langle N_{1}\rangle)/2\sigma^{2}$ is
plotted as a function of $Z$, here
$\sigma^{2}=(\sigma_{1}^{2}+\sigma_{2}^{2})/2$ was adopted, it
redisplays the $\alpha$ behavior in fig. \ref{fig:fig2}, its value
raises from the light fragments to heavy fragment, especially in
the heavy fragment part, it raises sharply. It decreases in the
region of light and intermediate mass fragments with increasing of
the reaction time, but opposite trend in the  region of heavy
fragments. This reveals that neglecting the second order in
equation (\ref{eq:seventeen}) is reasonable in most cases.

\begin{figure}
\includegraphics[width=0.5\textwidth]{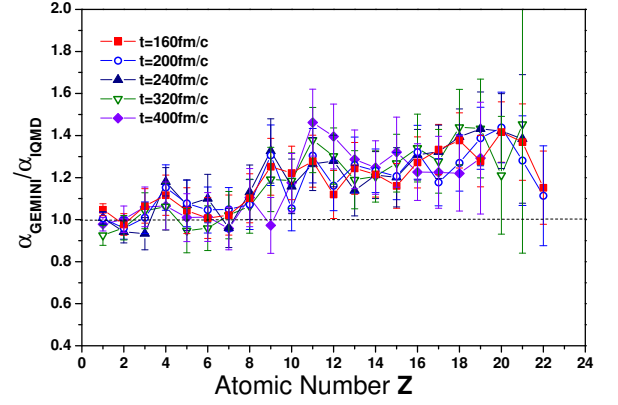}
\caption{\label{fig:fig4} The ratio of isoscaling parameters
$\alpha$ between final and primary products.}
\end{figure}

The secondary decay effect on isoscaling was investigated  by the
GEMINI code \cite{Ch88,Gemini} calculation. In this code, the
fragment charge number $Z$, fragment mass $A$ and the excitation
energy of the fragment is the only input for the GEMINI code. In
each event simulated by the IQMD, each fragment with charge number
great than 7 will consider sequential decay, the reason why
 the sequential decay for fragment charge number great
than 7 will be discussed later. Since in the IQMD model, the
angular momentum conservation was not taken into account, the
angular momentum of each fragment is not calculated, the angular
momentum of the hot product was set to zero as input to GEMINI.
Isoscaling behavior of final products after the secondary decay
are also observed. $\alpha$ parameters were extracted with
equation (\ref{eq:fourteen}) as a function of the fragment atomic
number $Z$ and shown in fig. \ref{fig:fig2}(b), one can find that
the dynamical effect due to different reaction times is still
clear for the light fragments, here the different reaction time
refers to the time the fragments were collected in IQMD
calculation and the GEMINI calculation starts. The secondary decay
effect, however, does not depend on the switching time of
dynamical IQMD to GEMINI code for the final products after
secondary decay. $\alpha$ basically keeps not change for the light
fragments even though the secondary decay, but  increases for the
heavy products with the increasing of the products atomic number
$Z$. In order to quantitatively compare the secondary decay effect
on isoscaling parameters, the ratio between $\alpha_{GEMINI}$ and
$\alpha_{IQMD}$ was plotted in fig. \ref{fig:fig3}, one can see
that  the ratios fluctuate around 1.0 for the light fragment with
atomic number $Z \leq 8$ and $\sim$ 1.2 for the products charge
number $ Z > 8$.

The light fragments include both the primary products which do not
experience sequential decay and the secondary decayed products
from the heavy primary products, however, this mixture of light
fragments do not affect the isoscaling phenomenon. It reveals that
the light primary fragments have same $N/Z$ as the final products
decayed from the heavy primary products, they have achieved local
equilibration or stable isotopic distribution on the isospin
degree of freedom, so $N/Z$ of light fragments is not affected by
the secondary decay process. It is a good observable to study the
primary products isospin information, which have been confirmed by
many experiment and theoretical
researches\cite{Xu00,Ts01,Ts01b,Br93,Bo02,Ge04,Sh04,On03,Ma04,Ti05}.

\begin{figure}
\includegraphics[width=0.5\textwidth]{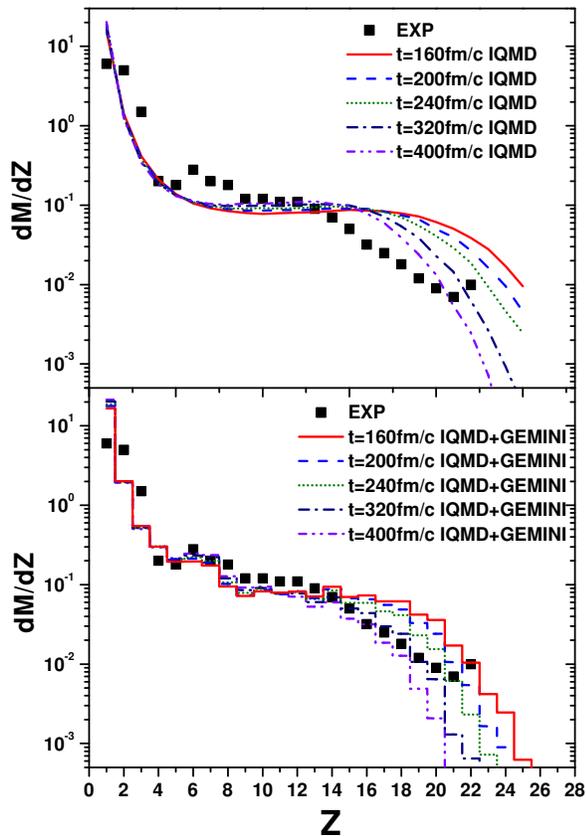}
\caption{\label{fig:fig5} Comparison of  charge distribution
between the experimental data (solid square) \cite{Wa98} and IQMD
calculation (line) or IQMD + GEMINI (histogram) calculation at
different reaction time.}
\end{figure}

In IQMD and GEMINI calculation, the fragment charge distributions
of IQMD and GEMINI with the experimental data \cite{Wa98} have
been compared in fig. \ref{fig:fig5}. The top panel is the
comparison of fragment charge distribution between experimental
data \cite{Wa98} and the IQMD calculation, it looks that heavier
products have been overestimated, but the intermediate mass
fragments were underestimated in calculations. This problem was
nicely solved after the secondary decay  with the GEMINI code was
taken  into account (see bottom panel of fig. \ref{fig:fig5}),
 especially when the switching time of IQMD fragments
 equals to 240 fm/$c$ or 320 fm/$c$. The best description
of the final charge distribution of products after GEMINI
calculation lies on the bump of $Z \approx 6$, this bump emerges
only when secondary decay of the primary fragments with $Z \geq 8$
are taken into account and while the secondary decay of primary
fragments with $Z < 8$ are not considered. This may be the
condition that the isoscaling parameters $\alpha$ is not affected
by the secondary decay.

\section{\label{sec:sec4}Density dependence of the symmetry energy coefficient}

\begin{table}
\caption{\label{tab:table2} Average excitation energies per
nucleon and the temperature calculated at different reaction times
for  two similar reaction systems $^{40}$Ca + $^{40}$Ca and
$^{48}$Ca + $^{48}$Ca  with incident energy $E_{inc}$ = 35 MeV/A
and impact parameter $b$ = 1fm.}

\begin{ruledtabular}
\begin{tabular}{llllll}
 $t$ &$160 fm/$c$$ &$200fm/$c$$ &$240fm/$c$$
 &$320fm/$c$$ &$400fm/$c$$ \\
\hline
$\langle E^{*}/A\rangle$\footnotemark[1] &4.812 &4.524 &4.274 &3.871 &3.584 \\
$T$\footnotemark[1] &6.937 &6.726 &6.537 &6.222 &5.987 \\
\hline
$\langle E^{*}/A\rangle$\footnotemark[2] &5.511 &5.214 &4.943 &4.496 &4.179 \\
$T$\footnotemark[2] &7.423 &7.221 &7.030 &6.705 &6.464 \\
\hline
$\langle T\rangle$\footnotemark[3] &7.180 &6.974 &6.784 &6.463 &6.226 \\
\end{tabular}
\end{ruledtabular}
\footnotetext[1]{average excitation energy per nucleon and
temperature in collision $^{40}$Ca + $^{40}$Ca }
\footnotetext[2]{average excitation energy per nucleon and
temperature in collision $^{48}$Ca + $^{48}$Ca}
\footnotetext[3]{average temperature $\langle T\rangle$ of
$^{40}$Ca + $^{40}$Ca and $^{48}$Ca + $^{48}$Ca reactions,
$\langle T\rangle=(T^{a}+T^{b})/2$}
\end{table}

The excitation energy per nucleon ($E^{*}/A$) was calculated from
equation (\ref{eq:thirteen}), $E^{*}/A$ distributions have the
Gaussian shape, from which the average excitation energy per
nucleon ($\langle E^{*}/A\rangle$) can be obtained. In table
\ref{tab:table2} list the average excitation energy per nucleon at
different reaction times for both similar reaction systems, the
temperature is extracted from equation $E^{*}/A = a \cdot T^{2}$,
$a = 1/10$ was adopted. The excitation energy of the system has
high values at early reaction time and it decreases with the
evolution of the system. At different reaction time the
temperature extracted from two reaction systems $^{40}$Ca +
$^{40}$Ca and $^{48}$Ca + $^{48}$Ca has approximately same values,
so the average temperature in this two reactions are calculated
simply to use as a parameter in following calculation, which are
listed in table \ref{tab:table2} too.

\begin{figure}
\includegraphics[width=0.5\textwidth]{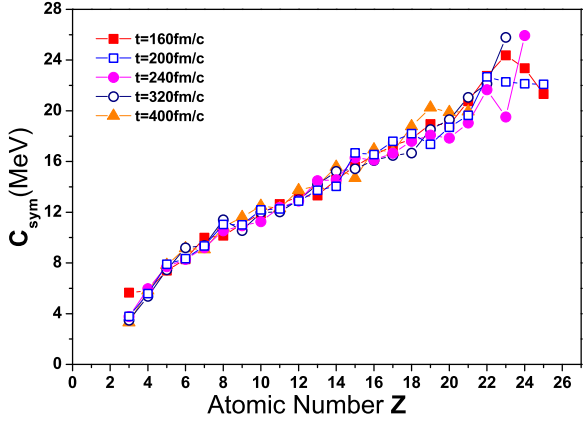}
\caption{\label{fig:fig6} Symmetry energy coefficients $C_{sym}$
as a function of the primary products atomic number $Z$ from IQMD
calculation.}
\end{figure}

\begin{figure}
\includegraphics[width=0.5\textwidth]{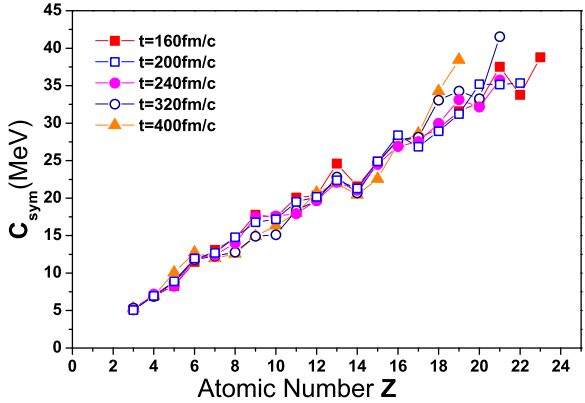}
\caption{\label{fig:fig7} Symmetry energy coefficients $C_{sym}$
as a function of the final products atomic number $Z$ from
IQMD+GEMINI calculation.}
\end{figure}

Using equation (\ref{eq:two}), from the extracted values of
$\alpha$, average temperature $\langle T\rangle$ of the two
systems, and the $\langle Z/A\rangle$ of the fragments in two
reaction systems, symmetry energy coefficient $C_{sym}$ in the EOS
can be derived. In present calculation, the $\langle Z/A\rangle$
was fitted from its $Z/A$ distributions of the primary and final
products respectively, which are Gaussian distribution too and can
get the average $Z/A$ in two reactions $\langle
Z_{1}/A_{1}\rangle$ and $\langle Z_{2}/A_{2}\rangle$. The derived
symmetry energy coefficient $C_{sym}$ was plotted as a function of
the atomic number $Z$ of products for primary products in fig.
\ref{fig:fig6} and final products in fig. \ref{fig:fig7}.

\begin{figure}
\includegraphics[width=0.5\textwidth]{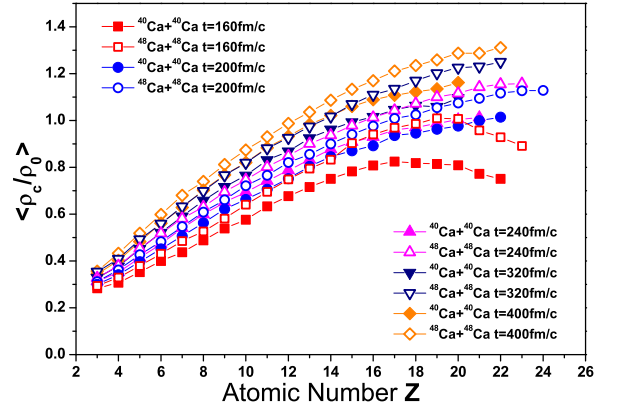}
\caption{\label{fig:fig8} Average  central density
$\langle\rho_{c}/\rho_{0}\rangle$ of the clusters in the rest
frame of cluster as a function of the cluster proton number $Z$ at
different times for two reactions.}
\end{figure}

One can find that the symmetry energy coefficient $C_{sym}$ for
different products with different atomic number $Z$ is not a
constant in the present calculation, it increases with the
increasing of the products charge number $Z$, and the dynamics
effect is almost vanished. To understand why the symmetry energy
coefficient $C_{sym}$ was not a constant for different products,
the densities of the products with different atomic number $Z$ in
IQMD simulation were extracted. In the IQMD frame, the fragments
are treated as cluster by simple coalescence model, the central
densities of clusters were calculated by equation (\ref{eq:four})
with $\vec{r} = 0$ in the cluster coordinate phase space, and sum
over all nucleons inside the same cluster. The central density
distribution $\rho_{c}/\rho_{0}$ of the clusters  is generally
Gaussian form and $\langle\rho_{c}/\rho_{0}\rangle$ can be fitted
from this Gaussian distribution, here $\rho_{0} = 0.16 fm^{-3}$
refers to the normal nuclear matter density. Fig. \ref{fig:fig8}
shows the fitted average central densities as a function of the
cluster charge number $Z$ at different reaction times in two
reactions $^{40}$Ca + $^{40}$Ca and $^{48}$Ca + $^{48}$Ca. The
light clusters are always in the very low density region, while
the heavy clusters can approach to the saturation density. Remind
that the nucleon can be considered to be emitted if its density is
lower than a certain low values, eg. $\rho_0/10$ in some transport
calculations, the above curve of $\rho(Z)$ support this argument.
For reaction system $^{48}$Ca + $^{48}$Ca, a little higher average
central densities $\langle\rho_{c}/\rho_{0}\rangle$ of clusters
than those in $^{40}$Ca + $^{40}$Ca are observed since the cluster
with same proton numbers $Z$ formed in reaction system $^{48}$Ca +
$^{48}$Ca has more neutrons than in reaction system $^{40}$Ca +
$^{40}$Ca, which has been verified by the $\langle N\rangle$
distribution from fig. \ref{fig:fig3}. The average
 central density of clusters in two reaction systems have some
divergence, in the later discussions, the average values of
$\langle \rho_{c}/\rho_{0}\rangle_{ave} = (\langle
\rho_{c}/\rho_{0}\rangle_{1}+\langle
\rho_{c}/\rho_{0}\rangle_{2})/2$ of these two reactions is
adopted.

\begin{figure}
\includegraphics[width=0.5\textwidth]{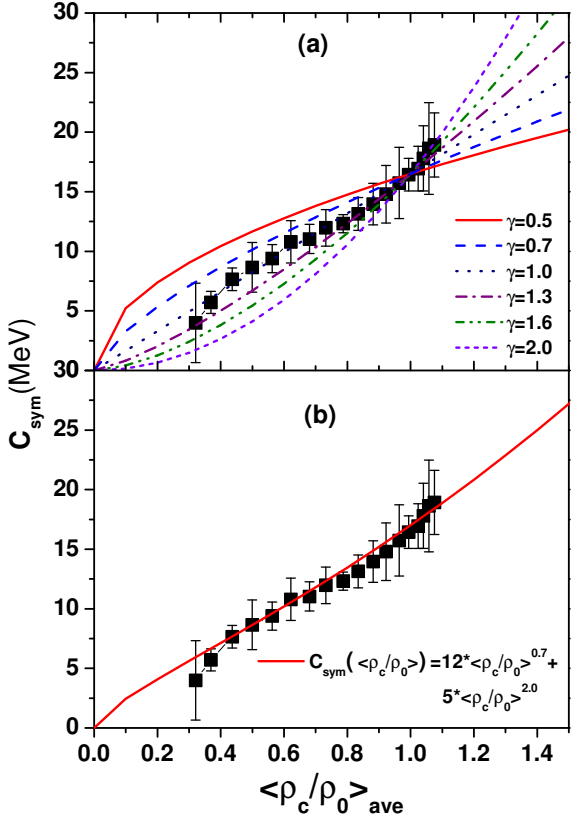}
\caption{\label{fig:fig9} The relationship between symmetry energy
coefficient $C_{sym}$ and the average central density
$\langle\rho_{c}/\rho_{0}\rangle_{ave}$, solid square represents
the IQMD calculations, the error bar represents the average over
different reaction times. Panel (a): different lines represent the
formula (\ref{eq:nineteen}) fitting with $C_{0} = 16.5$ MeV and
different parameter $\gamma$  which are shown in the insert; Panel
(b): Combination of two set parameters
$C_{sym}(\langle\rho_{c}/\rho_{0}\rangle) =
12\cdot\langle\rho_{c}/\rho_{0}\rangle^{0.7}
+5\cdot\langle\rho_{c}/\rho_{0}\rangle^{2.0}$}
\end{figure}

\begin{figure}
\includegraphics[width=0.5\textwidth]{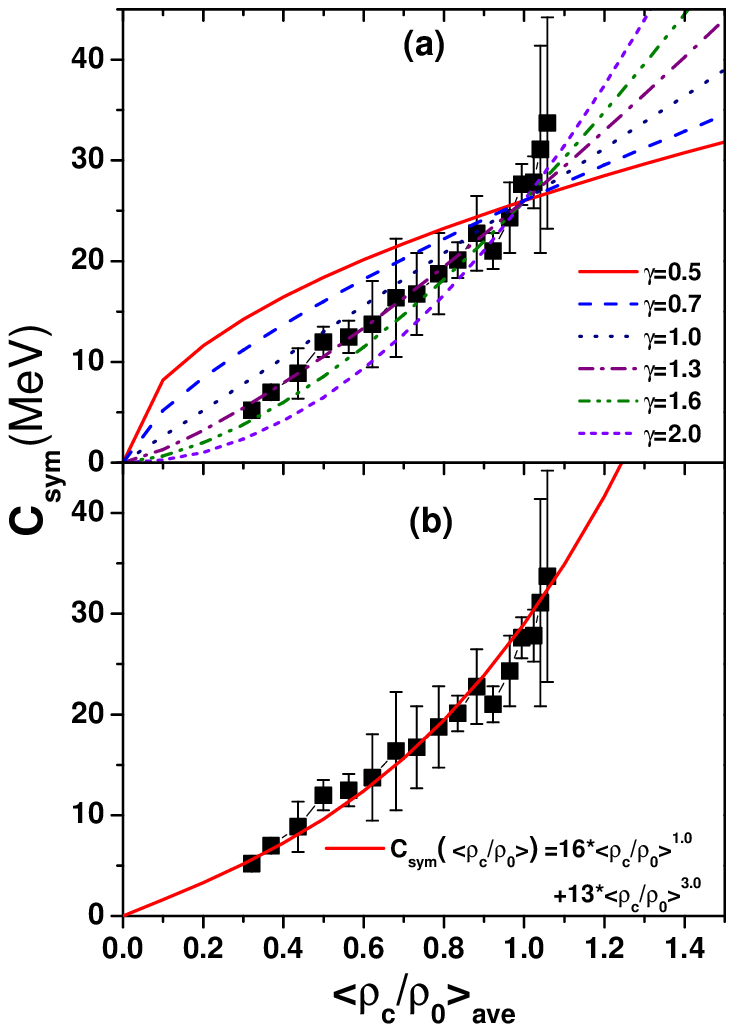}
\caption{\label{fig:fig10} The relationship between symmetry
energy coefficient $C_{sym}$ and the average central density
$\langle\rho_{c}/\rho_{0}\rangle_{ave}$, solid square represents
the IQMD+GEMINI calculations and the error bar represents the
average over different reaction times. Panel (a): different lines
represent the formula (\ref{eq:nineteen}) fitting with
$C_{0}=26MeV$ and different parameter $\gamma$ which are displayed
in the insert; Panel (b): Combination of two set parameters
$C_{sym}(\langle\rho_{c}/\rho_{0}\rangle)=
12\cdot\langle\rho_{c}/\rho_{0}\rangle^{0.7}
+5\cdot\langle\rho_{c}/\rho_{0}\rangle^{2.0}$}
\end{figure}

Since both symmetry energy coefficient $C_{sym}$ and the average
cluster central density $\langle\rho_{c}/\rho_{0}\rangle_{ave}$
are correlated with the  atomic number $Z$ of products, the
symmetry energy coefficient $C_{sym}$ can relate with the average
central density $\langle\rho_{c}/\rho_{0}\rangle_{ave}$ by
eliminating the products atomic number $Z$. The slight difference
of different reaction time information was expressed by adding
error bars to the derived $C_{sym}$. Fig.~\ref{fig:fig10}
demonstrates the relationship between symmetry energy coefficient
$C_{sym}$ and average central  density
$\langle\rho_{c}/\rho_{0}\rangle_{ave}$. The density dependence of
the symmetry energy term in equation of state (EOS) was generally
expressed by following equation
\begin{equation}
C_{sym}(\rho)=C_{0}\Big(\frac{\rho}{\rho_{0}}\Big)^{\gamma}
\label{eq:nineteen}
\end{equation}
where $\gamma$ represents the stiffness of the symmetry energy.
Some tests have been done for the determination of the parameter
$C_{0}$ and $\gamma$ in fig. \ref{fig:fig10}.  In panel (a) a
constant $C_{0}$ = 16.5 MeV is used but some different values of
$\gamma$, namely 0.5, 0.7, 1.0, 1.3, 1.6 and 2.0 were taken,
respectively, to fit the IQMD calculation results, one can find
that $\gamma$ = 1.0 seems to be the best fit to the IQMD
calculation than other $\gamma$ values in the low
$\langle\rho_{c}/\rho_{0}\rangle_{ave}<1.0$ region, but in
 $\langle\rho_{c}/\rho_{0}\rangle_{ave}>1.0$ region divergence arises between the IQMD
calculation and the fitting, it seems that in  high
$\langle\rho_{c}/\rho_{0}\rangle_{ave}>1.0$ part, $C_{sym}$ raises
sharply which corresponds to a $\gamma$ value great than 1.0. In
panel (b) the line is two different parameters combination of
equation (\ref{eq:nineteen}),
$C_{sym}(\langle\rho_{c}/\rho_{0}\rangle_{ave}) =
12\cdot\langle\rho_{c}/\rho_{0}\rangle_{ave}^{0.7}
+5\cdot\langle\rho_{c}/\rho_{0}\rangle_{ave}^{2.0}$, in the low
density region $\langle\rho_{c}/\rho_{0}\rangle_{ave}\leq1.0$, has
the similar shape with the above mentioned simple form
$C_{sym}(\langle\rho_{c}/\rho_{0}\rangle_{ave})
=16.5\cdot\langle\rho_{c}/\rho_{0}\rangle_{ave}^{1.0}$, but better
behavior than the simple form with only $\gamma = 1.0$. This
demonstrates that in the whole density region, the symmetry energy
coefficient $C_{sym}$ as a function of the nuclear density can not
be well described by only one simple equation (\ref{eq:nineteen})
in the present model calculation, but expressed by the various
contribution from two sets of EOS (\ref{eq:nineteen}). In present
calculation of IQMD, the simple equation such as
$C_{sym}(\langle\rho_{c}/\rho_{0}\rangle_{ave})  =
12\cdot\langle\rho_{c}/\rho_{0}\rangle_{ave}^{0.7} +
5\cdot\langle\rho_{c}/\rho_{0}\rangle_{ave}^{2.0}$ is a good
approximation for the $C_{sym}$ and
$\langle\rho_{c}/\rho_{0}\rangle_{ave}$.

Fig.~\ref{fig:fig10} plots the relationship between $C_{sym}$ and
$\langle\rho_{c}/\rho_{0}\rangle_{ave}$ after considering the
secondary decay. Panel (a) shows the IQMD + GEMINI calculation
$C_{sym}$ as a function of average central density
$\langle\rho_{c}/\rho_{0}\rangle_{ave}$, several fits with
 a constant $C_{0}$ = 26MeV and different $\gamma$ =
0.5, 0.7, 1.0, 1.3, 1.6, 2.0 of equation (\ref{eq:nineteen}) have
been made, in which $\gamma$ = 1.3 looks to fit the calculation
best in the density region
$\langle\rho_{c}/\rho_{0}\rangle_{ave}\leq1.0$, equation
(\ref{eq:nineteen}) with $\gamma > $1.3 works better in the higher
density part. Equation
$C_{sym}(\langle\rho_{c}/\rho_{0}\rangle_{ave})
=16\cdot\langle\rho_{c}/\rho_{0}\rangle_{ave}^{1.0}
+13\cdot\langle\rho_{c}/\rho_{0}\rangle_{ave}^{3.0}$
 was also adopted in panel (b) to fit the calculation in fig.
 \ref{fig:fig10}, it can give a  good coincidence
 between the calculation of IQMD+GEMINI and
the equation (\ref{eq:nineteen}) fitting.

No matter a simple form of equation (\ref{eq:nineteen}) was used,
or combination of different set parameters of equation
(\ref{eq:nineteen}) was used, different parameter $C_{0}$ was got
for the primary products in IQMD calculation and final products in
IQMD+GEMINI calculation. In the simple form of equation
(\ref{eq:nineteen}) for fitting IQMD calculation, $C_{0}$ = 16.5
MeV was got, while fitting IQMD + GEMINI calculation $C_{0}$ = 26
MeV was got. The different symmetry energy coefficients $C_{0}$
extracted from primary and final products indicate that the
secondary decay affects the extraction of symmetry energy
coefficient $C_{0}$.

From the fits to the IQMD and IQMD+GEMINI calculations, the
subnormal density region can be described better by a soft
symmetry energy with $\gamma<1.0$, the supernormal density region
can be described better by a stiff symmetry energy with
$\gamma>1.0$. This demonstrates that in the subnormal and
supernormal density regions, the symmetry energy shows different
behavior, the best description of the relationship between
symmetry energy coefficient $C_{sym}$ and the nuclear density
$\langle\rho/\rho_{0}\rangle$ in full density region should be
combination of equation (\ref{eq:nineteen}) with at least one
"soft" symmetry term with $\gamma \leq 1.0$ and one "stiff"
symmetry term with $\gamma
> 1.0$ \cite{Ch05}.

\section{\label{sec:sec5}Conclusions}

In summary, we applied the Isospin dependent Quantum Molecular
Dynamics (IQMD) model to investigate the isoscaling behavior in
 dynamical and statistical sequential decay
processes  of both reactions $^{40}$Ca + $^{40}$Ca and $^{48}$Ca +
$^{48}$Ca at $b$ = 1 fm. The calculation illustrates that both
primary and final isotopic yields  show the isoscaling behavior.

The isoscaling parameter $\alpha$ have similar values for the
light and intermediate mass fragments, but it increases much with
the fragments become heavy. $\alpha$ shows some dynamical effect,
i.e. it drops slowly with the increasing of reaction times. The
light final products which do not experience secondary decay
itself are not affected by the secondary decay, but the isoscaling
parameter $\alpha$ of the heavy products which experience
secondary decay process increase in comparison with the primary
products.

Average central density of the clusters is calculated in the IQMD
calculation, it  raises with the increasing of charge number of
clusters until the saturation density close to the normal nuclear
matter density $\rho_{0}$ = 0.16 fm$^{-3}$. From the equation
(\ref{eq:nineteen}), the density dependence of the symmetry energy
coefficient $C_{sym}$ can be extracted from the IQMD and
IQMD+GEMINI calculation. The results show that even though the
density dependence of the symmetry energy coefficient $C_{sym}$
can be approximately described by a $\gamma \sim 1.0$, it is more
reasonable to express this dependence by the combination of "soft"
and "stiff" symmetry energy term. From equation
(\ref{eq:nineteen}), the extracted symmetry energy coefficient
$C_{sym}$ is also affected by the secondary decay process, i.e. it
 leads to different values of $C_{0}$ and $\gamma$ from the initial values of
 IQMD.

\section{\label{sec:sec6}Acknowledgement}
This work was supported in part by the Shanghai Development
Foundation for Science and Technology under Grant Nos. 05XD14021
and 03QA14066, the National Natural Science Foundation of China
(NNSFC) under Grant Nos. 10535010, 10328259, 10135030, 10405032
and 10405033, and the Major State Basic Research Development
Program under Contract No. G200077404.


\begin{thebibliography}{99}
\bibitem{Li01}{\it Isospin Physics in Heavy Ion Collisions at
Intermadiate Energies}, edited by B.-A. Li and W. Schroeder(Nova
Science, New York, 2001)
\bibitem{To99}M. Di. Toro {\it et al}, Prog. Nucl. Nucl. Phys.
{\bf 42}, 125 (1999), {\it and references therein}
\bibitem{Li98}B.A. Li, C.M. Ko and W. Bauer, Int. J. Mod. Phys. E
{\bf 7}, 147 (1998), {\it and references therein}
\bibitem{Ma99}Y. G. Ma {\it et al}, Phys. Rev. C {\bf 60}, 024607 (1999)
\bibitem{Mu95}H. M\"{u}ller and B. D. Serot, Phys. Rev. C {\bf 52}, 2072 (1995)
\bibitem{Bo94}I. Bombaci, T. S. Kuo and U. Lombardo, Phys. Rep.
{\bf 242}, 165 (1994)
\bibitem{Gu01}S. Das Gupta, A. Z. Mekjian and M. B. Tsang, Adv. Nucl.
Phys. {\bf 26}, 91 (2001)
\bibitem{Xu00}H. S. Xu {\it et al},Phys. Rev. Lett. {\bf 85}, 716 (2000)
\bibitem{Ts01}M.B. Tsang {\it et al}, Phys. Rev. C {\bf 64},
041603(R) (2001)
\bibitem{Ts01a}M.B. Tsang {\it et al}, Phys. Rev. Lett. {\bf 86}, 5023
(2001)
\bibitem{Ts01b}M.B. Tsang {\it et al}, Phys. Rev. C. {\bf 64}, 054615
(2001)
\bibitem{Br93}J. Brzychczyk {\it et al}, Phys. Rev. C {\bf 47}, 1553
(1993)
\bibitem{Vo78}V. Volkov, Phys. Rep. {\bf 44}, 93 (1978)
\bibitem{Ve04}M. Veselsky, G.A. Souliotis and M. Jandel, Phys.
Rev.  C {\bf 69}, 044607 (2004)
\bibitem{So03}G.A. Souliotis {\it et al}, Phys. Rev. C {\bf 68},
024605 (2003)
\bibitem{Bo02}A.S.Botvina, O.V. Lozhkin and W. Trautmann, Phys.
Rev. C {\bf 65}, 044610 (2002)
\bibitem{Ge04}E. Geraci {\it et al}, Nucl Phys. A {\bf 732}, 173 (2004)
\bibitem{Sh04}D.V.Shetty {\it et al}, Phys. Rev. C {\bf 70},
011601(R) (2004)
\bibitem{Wa05}K. Wang, Y. G. Ma, Y. B. Wei, X. Z. Cai, J. G.
Chen, D. Q. Fang,W. Guo, G. L. Ma,W. Q. Shen,W. D. Tian, C. Zhong,
and X. F. Zhou, Chin. Phys. Lett. 22, 53 (2005)
\bibitem{Ma05}Y. G. Ma, K. Wang, X. Z. Cai, J. G. Chen, J. H. Chen, D. Q. Fang, W. Guo, C. W. Ma, G. L. Ma, W. Q. Shen,
Q. M. Su, W. D. Tian, Y. B. Wei, T. Z. Yan, C. Zhong, X. F. Zhou,
and J. X. Zuo, Phys. Rev. C {\bf 72} 064603 (2005)
\bibitem{Ma04}Y. G.Ma, K. Wang, Y. B.Wei, G. L.Ma, X. Z. Cai, J. G.
Chen, D. Q. Fang, W. Guo, W. Q. Shen, W. D. Tian, and C. Zhong,
Phys. Rev. C 69, 064610 (2004).
\bibitem{On03}A. Ono, P. Danielewicz, W. A. Friedman, W. G. Lynch, and
M. B. Tsang, Phys. Rev. C {\bf 68}, 051601(R) (2003)
\bibitem{Ti05}W. D. Tian, Y. G. Ma, X. Z. Cai, J. G. Chen, J. H. Chen,
D. Q. Fang,W. Guo, C.W. Ma, G. L.Ma,W. Q. Shen, K.Wang, Y. B.Wei,
T. Z. Yan, C. Zhong, and J. X. Zuo, Chin. Phys. Lett. 22, 306
(2005).
\bibitem{Ra05}Ad. R. Raduta, Eur. Phys. J. A {\bf 24}, 85 (2005)
\bibitem{La00}J. M. Lattimer and M. Prakash, Phys. Rep. {\bf 333},
121 (2000)
\bibitem{My66}W. D. Myers and W.J. Swiatecki, Nucl. Phys. A {\bf 81}, 1
(1966)
\bibitem{Fe05}A. Le F\`{e}vre {\it et al}, Phys. Rev.
Lett. {\bf 94}, 162701 (2005)
\bibitem{Ch05}Lie-Wen Chen, Che Ming Ko and Bao-An Li, Phys. Rev.
Lett. {\bf 94}, 032701 (2005)
\bibitem{Ai88-91}J. Aichelin {\it et al}, Phys. Rev. C {\bf 37}, 2451
(1988); J. Aichelin, Phys. Rep. {\bf 202} 233 (1991)
\bibitem{Ch98}Chen Liewen, Zhang Fengshou and Jin Genming, Phys.
Rev. C {\bf 58}, 2283 (1998)
\bibitem{On92}A. Ono, H. Horiuchi, T. Maruyama, A. Ohnishi, Prog.
Theo. Phys. {\bf 87}, 1185(1992)
\bibitem{Pa01}M. Papa, T. Maruyama, A. Bonasera, Phys. Rev. C {\bf
64}, 024612 (2001)
\bibitem{Ch68}K. Chen {\it et al}, Phys. Rev. {\bf 166}, 949
(1968)
\bibitem{Ch88}R.J. Charity {\it et al}, Nucl. Phys. A {\bf 483}, 371
(1988)
\bibitem{Gemini}R.J. Charity, computer code GEMINI, see http:// wunmr.wustl.edu/pub/gemini
\bibitem{Wa98}R. Wada {\it et al}, Phys. Lett. B {\bf 422}, 6
(1998)
\end{thebibliography}

\end{document}